\documentclass[preprint,3p,times]{elsarticle}

\usepackage{amssymb}
\usepackage{amsthm}
\usepackage{amsmath}
\usepackage{epsfig}
\usepackage{subfigure}
\usepackage[usenames]{color}
\usepackage{algorithm}
\usepackage{algorithmic}
\usepackage{nicefrac}
\usepackage{caption}
\epsfverbosetrue
\graphicspath{{Figs/}{../Figs/}}

\newcommand{\mb}[1]{\ensuremath{\mathbf{#1}}}

\journal{International Journal for Numerical Methods in Engineering}

\begin{document}
 \setlength{\parindent}{0.0ex}
 \setcounter{secnumdepth}{4}
 \setcounter{tocdepth}{4}
\begin{frontmatter}



\title{Extended Comment on the Article "Consistent Development of a Beam-To-Beam Contact Algorithm via the Curve to Solid Beam Contact - Analysis for the Non-Frictional Case"}


\author[lnm]{Christoph Meier\corref{cor1}}
\ead{meier@lnm.mw.tum.de}
\author[ubm]{Alexander Popp}
\author[lnm]{Wolfgang A. Wall}

\address[lnm]{Institute for Computational Mechanics, Technical University of Munich, Boltzmannstrasse 15, 85748 Garching b. M{\"u}nchen, Germany}
\address[ubm]{Institute for Mathematics and Computer-Based Simulation, University of the Bundeswehr Munich, Werner-Heisenberg-Weg 39, 85577 Neubiberg, Germany}

\cortext[cor1]{Corresponding authors}
\end{frontmatter}

%
\section{Motivation for this comment}
\label{introduction}
%

In the authors' previous work~\cite{meier2015b,meier2015c}, novel finite element formulations for the contact interaction of slender beams with undeformable circular cross-sections have been proposed, which consider contact interaction either via a point force model (denoted as "point-to-point contact" in~\cite{meier2015b}), a line force model (denoted as "line-to-line contact" in~\cite{meier2015b}) or a combined point and line force model (denoted as "all-angle beam contact [ABC]" in~\cite{meier2015c}). In this context, the authors also analyze the uniqueness and existence of closest point projections (CPP) between the contacting beams, either for the so-called \textit{bilateral} CPP (yielding one pair of closest points between the two beam \textit{centerline} curves, as required for the point-to-point contact model) or for the so-called \textit{unilateral} CPP (yielding the closest point on the \textit{centerline} of a second / master beam for any given point on the \textit{centerline} of the first / slave beam, as required for the line-to-line contact model). In their recent article~\cite{konyukhov2017}, Konyukhov et al. propose an algorithm for beam-to-beam contact based on the so-called curve-to-solid beam (CTSB) contact model. The basic idea to avoid the bilateral closest point projection underlying the point-to-point contact model (which is not unique for certain geometrical configurations) by employing a line force contact model is similar to the line-to-line contact formulation, extended, however, by a model for cross-section deformation of the master beam (the associated CPP yields the closest point on the \textit{surface} of the master beam for any given point on the \textit{centerline} of the slave beam). In this article, Konyukhov et al. refer extensively to the aforementioned works~\cite{meier2015b,meier2015c} by the authors. However, many of these statements turn out to be scientifically incorrect and not only question the quality and correctness of the works~\cite{meier2015b,meier2015c} in a way that is neither objective nor justified but also might cause quite some confusion to researchers in this field. Hence, the authors find it necessary to comment on these statements, disprove them if incorrect, and demonstrate the correctness of the derivations made in~\cite{meier2015b,meier2015c}.\\

More specifically, in their recent contribution~\cite{meier2015b}, the authors proposed a reformulation of the general criteria for existence and uniqueness of the bilateral CPP on the basis of mathematically consistent inequality estimates in order to end up with more compact, alternative criteria eventually allowing for simpler and more efficient numerical formulations. Konyukhov et al.~\cite{konyukhov2017} argue that this reformulation is not admissible from a mathematical point of view, and that the resulting alternative criteria as well as the numerical formulations derived on the basis of these criteria are incorrect and not applicable to general beam contact problems. Moreover, Konyukhov et al.~\cite{konyukhov2017} claim to verify these statements by means of numerical examples, where the alternative criteria derived by the authors~\cite{meier2015b} reputedly fail. In fact, however, these statements by Konyukhov et al. are not correct, i.e. all of the numerical examples presented by Konyukhov et al.~\cite{konyukhov2017} can correctly be captured by the criteria derived in~\cite{meier2015b}. Moreover, it can be proven in a straight-forward manner that the reformulation proposed by the authors is mathematically concise and applicable to arbitrary beam contact problems as long as the underlying requirements formulated in~\cite{meier2015b} are fulfilled. In the following, this central point of criticism by Konyukhov et al. will be discussed in detail, not least to prevent a potential impediment of future developments and scientific progress in this field of research. Finally, it has to be pointed out that the contribution of Konyukhov et al.~\cite{konyukhov2017} contains quite a number of further incorrect or misleading statements. For completeness, and as it might be helpful for researchers in this field, these statements along with a corresponding response are summarized in~\ref{further_criticism} of this article.

%
\section{Central point of criticism in the referenced article}
\label{central_criticism}
%

In Remark 3 (pages 18 and 19 in~\cite{konyukhov2017}), Konyukhov et al. state that the general criteria for existence and uniqueness of the bilateral CPP "are a system of inequalities determining the geometrical region of existence and uniqueness", which "cannot be estimated" in a manner as done by the authors~\cite{meier2015b}. Moreover, Konyukhov et al. claim to show in their article "that only the full criteria ... must be used to judge the existence and uniqueness of the curve-to-curve CPP procedure" while the estimates made by the authors in~\cite{meier2015b} "cannot be used at all". In fact, Konyukhov et al. do not provide any kind of proof or verification for this statement, i.e. an example where the estimates in~\cite{meier2015b} actually do fail although the underlying assumptions are met. In the following, the correctness of the estimates made by the authors in~\cite{meier2015b} will be demonstrated. Interestingly, it can be shown (see~\ref{anhang:importance_of_assumptions}) that the application of the curve-to-solid beam contact scheme proposed by Konyukhov et al. to general beam contact problems (involving curved beam elements with arbitrary mutual orientation) requires the same additional assumptions as the estimates made by the authors in~\cite{meier2015b}.\\

For later use, the general criteria for existence and uniqueness of the bilateral CPP, i.e. equations (25) and (32) in the original contribution of the authors~\cite{meier2015b}, shall briefly be restated here:
\begin{subequations}
\label{criterion_general}
\begin{align}
\bar{\kappa}_{2}  d \cos(\beta_2)\dot{<}1,\label{criterion_general_a}\\
\big(1 + \bar{\kappa}_1 d \cos(\beta_1)\big) \big( 1-\bar{\kappa}_2 d \cos(\beta_2) \big) \dot{>} \cos(\alpha)^2. \label{criterion_general_b}
\end{align}
\end{subequations}
Here, $d$ is the distance between the pair of closest points on the beam centerlines, $\alpha  \in [0;90^{\circ}]$ is the angle between the centerline tangent vectors at the contact points, $\bar{\kappa}_1$ and $\bar{\kappa}_2$ are the centerline curvatures at the contact points, and $\beta_1$ as well as $\beta_2$ represent the angles between the contact normal vector and the respective normal vectors of the Frenet-Serret triads associated with the beam centerlines at the contact points. In~\cite{konyukhov2017}, the same quantities are denoted as $r$, $\psi$, $k_1$, $k_2$, $\pi-\varphi_1$ and $\varphi_2$, respectively. Moreover, the point-to-point and line-to-line contact model are denoted as curve-to-curve and point-to-curve contact model in~\cite{konyukhov2017}. Throughout this document, the notation used in~\cite{meier2015b} will be employed. The general criteria~\eqref{criterion_general} (i.e. equations (25) and (32) in the authors' work~\cite{meier2015b}) are equivalent to the expressions in equation (63) by Konyukhov et al.~\cite{konyukhov2017} (as already derived in~\cite{konyukhov2010}). However, in the authors' work~\cite{meier2015b}, these criteria only represent an intermediate result. Specifically, two additional assumptions are made, which enable a reformulation - and a considerable simplification - of these criteria as discussed below.\\

\hspace{0.6cm}
\begin{minipage}{15.0 cm}
\textit{Remark - Alternative derivation of general criteria and resulting notation:} In the authors' contribution~\cite{meier2015b}, the general criteria~\eqref{criterion_general} are derived (in a different manner than in~\cite{konyukhov2010}) by defining the bilateral CPP as extremum of the distance function associated with the unilateral CPP. In this context, the subscript $1$ refers to the slave beam and the subscript $2$ to the master beam of the unilateral CPP. Moreover, criterion~\eqref{criterion_general_a} ensures existence and uniqueness of the underlying unilateral CPP. For existence and uniqueness of the bilateral CPP, criterion~\eqref{criterion_general_b} has to be fulfilled in addition to~\eqref{criterion_general_a}.\\
\end{minipage}

In fact, the criteria~\eqref{criterion_general} represent a system of inequalities, and the reformulation of these inequalities by means of proper inequality estimates, i.e. by employing upper / lower bounds as done in~\cite{meier2015b} is a standard mathematical procedure. There is no reason, neither from a mathematical nor from an engineering point of view, why such an approach would be incorrect. In the following, the main steps of this reformulation shall be recapitulated. Thereto, beams with undeformable circular cross-sections (both of radius $R$) are considered, as well as the following two assumptions:
\renewcommand{\labelenumi}{\roman{enumi})}
\begin{enumerate}
\item The potential contact partners are already sufficiently close: $d \leq 2R$.
\item The centerline curvature $\bar{\kappa}$ is small compared to the cross-section radius $R$: $\mu_{max} := R max(\bar{\kappa}) < 0.5$.
\end{enumerate}
Assumption i) is reasonable since CPPs only have to be evaluated if contact / penetration actually occurs, i.e. if the gap $g:=d-2R$ is negative, or in other words, if $d \leq 2R$. In this context, assumption i) can easily be reformulated to $d \leq k\cdot 2R$ with a reasonably small safety factor $k > 1$ in order to guarantee for unique CPPs already in the range of small positive gaps (which is e.g. required to check a pair of close beams, as typically provided by a contact search algorithm, for active contacts / negative gaps). Also the consideration of beams with different cross-section radii (as long as those are of the same order of magnitude) is straight-forward by replacing the expression $2R$ with $R_1+R_2$. The general results derived in~\cite{meier2015b} can easily be adapted to these slight modifications. For simplicity, the original assumption i) as stated above will be considered in the following. Assumptions similar to ii) are absolutely standard for geometrically nonlinear (structural) beam models: In order to limit the model error between 1D beam theory and 3D contiuum theory to an acceptable level, such beam theories typically only allow for small centerline curvatures compared to the cross-section radius, i.e. for a maximal admissible ratio $\mu_{max}:= R max(\bar{\kappa}) \ll 1$ (see e.g. Geradin and Cardona~\cite{geradin2001}, Linn et al.~\cite{linn2013} or Meier et al.~\cite{meier2016}; if beams with different cross-section radii $R_1$ and $R_2$ shall be considered, these radii $R_i$ should at least be of the same order of magnitude in order to still fulfill the requirement $\mu_{max}= max(R_i) max(\bar{\kappa}) \ll 1$). Consequently, also this second assumption does typically not represent an additional limitation concerning the range of potential / suitable applications. If assumptions i) and ii) are valid, and by taking advantage of the extreme values $\pm1$ of the cosine function, it is a standard mathematical procedure to reformulate the criteria~\eqref{criterion_general} by means of inequality estimates. As a result, it is found that~\eqref{criterion_general} is always fulfilled, i.e. a unique CPP solution exists, if the following inequalities hold (which are identical to equations (26) and (34) in~\cite{meier2015b}):\\
 \begin{subequations}
\label{criterion_estimate}
\begin{align}
2 \mu_{max} \dot{<}1, \label{criterion_estimate_a}\\
\left(1 - 2 \mu_{max} \right)^2 \dot{>} \cos(\alpha)^2 \quad \rightarrow \quad \alpha \dot{>} \alpha_{min}=\arccos \left(1 - 2 \mu_{max} \right). \label{criterion_estimate_b}
\end{align}
\end{subequations}
Of course, as consequence of the employed estimates, there are cases where the inequalities~\eqref{criterion_estimate} are not fulfilled, but a unique CPP solution exists nonetheless (i.e. criteria~\eqref{criterion_general} are still true). Mathematically, criteria~\eqref{criterion_estimate} are a sufficient, but not a necessary condition for the fulfillment of criteria~\eqref{criterion_general}. For that reason, the criteria~\eqref{criterion_estimate} are denoted as worst-case or conservative estimates, or in other words: An interval of contact angles $\alpha  \in ]\alpha_{min};90^{\circ}]$ with unique bilateral CPP can correctly be predicted, even though that may not be the largest possible interval with unique bilateral CPP. Correspondingly, $\alpha_{min}$ is a conservative lower bound, but not the smallest possible lower bound. While the criteria~\eqref{criterion_general} consider CPPs between arbitrary space curves, the criteria~\eqref{criterion_estimate} take advantage of additional properties (reflected by assumptions i) and ii)) that are specific for beam contact problems.\\

\hspace{0.6cm}
\begin{minipage}{15.0 cm}
\textit{Remark - Benefits arising from simplified criteria:} Two main conclusions can be drawn from the criteria~\eqref{criterion_estimate}, as long as assumptions i) and ii) are valid: First, the criterion~\eqref{criterion_estimate_a} is always fulfilled due to assumption ii), i.e. a unique solution of the unilateral CPP solution exists (and the associated line-to-line contact model can be applied) for arbitrary configurations of the contacting beams (i.e. arbitrary mutual orientations as well as closest point distances and curvatures fulfilling i) and ii)). Second, the criterion~\eqref{criterion_estimate_b} required for a unique bilateral CPP could be considerably simplified and expressed by only one deformation-dependent kinematic quantity, namely the contact angle $\alpha$ (while the ratio $\mu_{max}$ is an a priori known / constant parameter). The potential advantages of the simplified criteria~\eqref{criterion_estimate} are obvious: The ultimate goal of the authors' works~\cite{meier2015b,meier2015c} is to derive a contact formulation that switches smoothly (and variationally consistently) from the point-to-point to the line-to-line contact model for configurations where no unique bilateral CPP exists, i.e. where the point-to-point contact model cannot be applied. Such a smooth model transition would be rather complex, the resulting finite element formulations computationally expensive, and the nonlinear solution scheme probably less robust if it relied on the criteria~\eqref{criterion_general} with six displacement-dependent kinematic quantities that would have to be considered in consistent variation and linearization procedures. Moreover, also the geometrical interpretation of the criteria~\eqref{criterion_estimate} is straight-forward, while a  direct analysis of the general criteria~\eqref{criterion_general} "is rather complicated" as even stated by Konyukhov et al.~\cite{konyukhov2010}.\\
\end{minipage}

%
\section{Summary and Conclusion}
\label{introduction}
%

It can be concluded that the criteria for existence and uniqueness of closest point projections in beam-to-beam contact problems as derived by Meier et al. are correct from a mathematical point of view and reasonable from a mechanical point of view. Specifically, the reformulation of the general criteria~\eqref{criterion_general} to the simplified criteria~\eqref{criterion_estimate} as proposed by the authors in~\cite{meier2015b} is mathematically consistent, which is in strong contrast to the statements made by Konyukhov et al.~\cite{konyukhov2017}. The predicted scope of configurations with unique bilateral closest point projections is correct provided that requirements i) and ii) stated in~\cite{meier2015b}, which are standard for typical beam contact models and mechanical beam theories, are fulfilled. For all the examples with non-unique bilateral closest point projection as presented by Konyukhov et al.~\cite{konyukhov2017}, it can easily be verified that also the criteria~\eqref{criterion_estimate} proposed by the authors will correctly identify the non-uniqueness of the closest point projections, i.e. also these criteria  will not guarantee for a unique closest point projection since at least one of the assumptions i) and ii) is violated. Thus, in contrast to their claims, Konyukhov et al.~\cite{konyukhov2017} cannot disprove - neither analytically nor numerically - the mathematical correctness of the criteria proposed by the authors.

\appendix

%
\section{Importance of additional assumptions for other types of beam contact models}
\label{anhang:importance_of_assumptions}
%

So far, it has been demonstrated that assumptions i) and ii) enable a mathematically consistent reformulation / simplification of the extistence and uniqueness criteria for the bilateral CPP underlying the point-to-point contact model, as proposed by the authors~\cite{meier2015b}. In the following two paragraphs, it will be shown that assumptions comparable to i) and ii) are not unusual and typically also required for other types of beam-to-beam contact models such as the line-to-line contact model or even for the curve-to-solid beam contact model proposed by Konyukhov et al.~\cite{konyukhov2017}. For example, the general criterion for existence of the unilateral CPP underlying the line-to-line contact model is given by~\eqref{criterion_general_a}. For a configuration with given angle $\beta_2$ and closest point distance $d$, criterion~\eqref{criterion_general_a} states a requirement for the maximally admissible centerline curvature $\bar{\kappa}_{2}$ of the master beam. If the line-to-line contact model shall be capable of representing arbitrary contact configurations (contact of curved beams with arbitrary mutual orientations), criterion~\eqref{criterion_general_a} has to be fulfilled for arbitrary values of $\beta_2$. Since $\bar{\kappa}_2>0$ and $d>0$ hold per definition, the strictest requirement, i.e. the smallest value for the maximally admissible centerline curvature $\bar{\kappa}_{2}$, results for configurations with $\beta_2=0$ (contact normal vector and Frenet-Serret normal vector of the master beam are parallel). For such configurations, a non-unique unilateral CPP can occur if $\bar{\kappa}_{2}  \geq 1/d$ (see~\eqref{criterion_general_a} for $\beta_2=0$), e.g. if the closest point of the slave beam is located at the center of curvature of the master beam centerline evaluated at the corresponding closest point (see e.g. Figure~\ref{fig:limitations_circle}, where $\bar{r}_{2}:=1/\bar{\kappa}_{2}=d$). Thus, only for configurations fulfilling the requirement $\bar{\kappa}_{2}  < 1/d$, a unique unilateral CPP can be guaranteed (independent of $\beta_2$, i.e. independent of the mutual orientations of the beams). In order to release the restriction of small curvatures $\bar{\kappa}_{2}  < 1/d$ as far as possible, it is reasonable to only consider sufficiently small distances $d$, i.e. distances $d \leq 2R$ (see assumption i), where contact can actually occur. Under this assumption, a unique unilateral CPP for beams with arbitrary mutual orientations (i.e. arbitrary $\beta_2$) can only be guaranteed, if $\bar{\kappa}_{2}  < 1/(2R) \rightarrow \mu_{max} := R max(\bar{\kappa}) < 0.5$, which is identical to assumption ii). Thus, also the well-known line-to-line contact model (denoted as point-to-curve contact model by Konyukhov et al.~\cite{konyukhov2017}) requires the fulfillment of assumptions i) and ii), if it shall be applied to general contact problems involving beams with arbitrary mutual orientations.\\

In the following, the discussion carried on for the line-to-line contact model above shall be extended to the curve-to-solid beam contact model proposed by Konyukhov et al.~\cite{konyukhov2017}.  In the curve-to-solid beam contact model, the slave beam is modeled similar to the line-to-line contact model, i.e. as beam with constant / rigid circular cross-section shape, while the cross-section of the master beam is assumed to be deformable. For a given integration point on the slave beam centerline, the associated closest point on the master beam is found iteratively via projection onto the master beam \textit{surface}, yielding two closest point surface coordinates (while the line-to-line contact model is based on a projection of the same integration point onto the master beam \textit{centerline}). It is important to notice that for configurations where the cross-sections of the master beam are circular with constant cross-section radius, the resulting CPPs of the curve-to-solid beam model and of the line-to-line contact model are identical (since a contact normal vector that is perpendicular to the centerline of a cylinder segment is also perpendicular to the cylinder surface at the penetration point). For such configurations (circular master beam cross-section with constant radius) and scenarios where no unique (unilateral) CPP of the line-to-line contact model exists (see e.g. Figure ~\ref{fig:limitations_circle}, with the straight beam representing the slave), also the CPP of the curve-to-solid beam contact model is not unique. In other words, the application of the curve-to-solid beam contact model to general contact problems involving beams with arbitrary mutual orientations requires the same assumptions concerning admissible distances / curvatures of the contacting beams as stated for the line-to-line contact model in the paragraph above. These additional restrictions were not stated in~\cite{konyukhov2017} (see the discussion of existence and uniqueness of the CPP underlying the curve-to-solid beam contact model in Section 5.2 of~\cite{konyukhov2017}) since only straight / first-order beam elements had been considered by Konyukhov et al. For straight beam elements, however, the analysis is considerably simplified. In this case, existence and uniqueness e.g. of the unilateral CPP underlying the line-to-line contact model can be guaranteed for arbitrary configurations (without any additional restrictions / assumptions such as i) and ii)). Even the general existence and uniqueness criteria~\eqref{criterion_general} for the bilateral CPP underlying the point-to-point contact model simplify to the trivial requirement $\alpha>0$ for straight / first-order beam elements. However, once higher-order beam elements are considered (e.g. based on higher-order Hermite / Lagrange polynomials or on isogeometric approaches as stated in~\cite{konyukhov2017}), also the curve-to-solid beam contact model proposed by Konyukhov et al.~\cite{konyukhov2017} (if applied to general beam contact problems involving curved beams with arbitrary mutual orientation), requires the additional assumptions i) and ii), even though not mentioned explicitly in~\cite{konyukhov2017}. This is an interesting insight, given that the estimates made by the authors~\cite{meier2015b} on the basis of assumptions i) and ii) were one major source for criticism stated by Konyukhov et al.~\cite{konyukhov2017}.

%
\section{Further points of criticism in the referenced article}
\label{further_criticism}
%

As already mentioned above, the contribution of Konyukhov et al.~\cite{konyukhov2017} contains quite a number of additional wrong or misleading statements, which, for the sake of brevity, however, have not been discussed above. For completeness, these statements along with a corresponding response are summarized in the following.

%
\subsection{Statement 1}
\label{statement_1}
%

Also in Remark 3 (pages 18 and 19 in~\cite{konyukhov2017}), Konyukhov et al. argue that the estimates employed by the authors~\cite{meier2015b} would not be admissible since "the multiplicity of the CPP solution" (equation (62) in~\cite{konyukhov2017}) "does not depend either on a cross-section radius $R$ for a beam, or on an angle between tangent vectors $\psi$." This argumentation is not correct since the mentioned criterion in equation (62) of~\cite{konyukhov2017} (which is equivalent to~\eqref{criterion_general_b} above) clearly depends on the angle between the tangent vectors $\psi$ (or $\alpha$ in the notation used here). Moreover, the dependence on the cross-section radius $R$ in the criterion~\eqref{criterion_estimate_b} (via $\mu_{max}$) is a mere consequence of re-expressing the closest point distance $d$ in~\eqref{criterion_general_b} as multiple of the cross-section radius (e.g. $d \leq 2R$ in case of contact, see assumption i)).

%
\subsection{Statement 2}
\label{statement_2}
%

At the end of Remark 3 (pages 18 and 19 in~\cite{konyukhov2017}), Konyukhov et al. claim that the estimates made in~\cite{meier2015b} employ $\cos \beta_1 \approx -1$ and $\cos \beta_2 \approx 1$, and, thus, that theses estimates would "implicitly assume" or "enforce geometrically" the equality of the Frenet-Serret normal vectors of the two beams at the contact point, i.e. $\bar{\mb{n}}_1 \approx \bar{\mb{n}}_2$. Also this statement has to be contradicted: As shown above, the estimates made in the reformulation from~\eqref{criterion_general} to~\eqref{criterion_estimate} take advantage of the extreme values $\pm 1$ of the cosine function, but this is by no means equivalent to geometrically enforcing configurations with $\cos \beta =  \pm 1$. The geometrical interpretation is rather as follows: In scenarios where mechanical equilibrium configurations are characterized by $\cos \beta =  \pm 1$, the worst case estimates in~\eqref{criterion_estimate} are identical to the corresponding terms in the general criteria~\eqref{criterion_general} (see e.g. the "double helix" example in Section 6.2 of ~\cite{meier2015b}, where both criteria~\eqref{criterion_general} and~\eqref{criterion_estimate} yield an identical lower bound $\alpha_{min}$ for the contact angle).

%
\subsection{Statement 3}
\label{statement_3}
%

At the beginning of Section 7.2, Konyukhov et al.~\cite{konyukhov2017} claim that the variationally consistent transition between point-to-point contact and line-to-line contact as proposed in~\cite{meier2015c} (denoted as ABC formulation) would be based "on a mathematically incorrect estimation" and that the numerical examples in this work and the subsequent contribution~\cite{meier2016b} would only employ the line-to-line contact alogrithm - and not the ABC formulation - "to treat this on the safe side". First, the correctness of the estimates made in the derivation of~\eqref{criterion_estimate} has already been discussed above. The underlying details shall not be repeated here. Second, it can easily be checked that the majority of the numerical examples in~\cite{meier2015c} and~\cite{meier2016b} are actually based on the proposed ABC formulation (with smooth model transition between point-to-point and line-to-line contact). Only for examples where the resulting range of contact angles was expected to lie below $\alpha_{min}$ according to~\eqref{criterion_estimate_b} (i.e. in a regime where the ABC formulation would select a line-to-line contact model anyway), a pure line-to-line contact formulation has been applied a priori. Of course, the result would have been the same if the ABC formulation in combination with a properly prescribed transition range had been applied.

%
\subsection{Statement 4}
\label{statement_4}
%

In Section 5.4.3, Konyukhov et al.~\cite{konyukhov2017} consider a special class of space curves, denoted as "parallel curves". Specifically, for an arbitrary curve $\mb{r}_a(s)$, with $s$ representing an arc-length parameter of this curve, a "parallel curve" $\mb{r}_b(s)$ (parametrized by the same parameter $s$) is constructed via $\mb{r}_b(s):=\mb{r}_a(s)+d_0 \mb{e}_a(s)$, where $\mb{e}_a(s)$ represents a prescribed unit normal vector field associated with the curve $\mb{r}_a(s)$, and the closest point distance $d_0$ is constant along the curves (here, the two curves are identified by indices $a$ and $b$ to avoid confusion with the indices $1$ and $2$ used above to distinguish the slave and master beam). As consequence of the constant distance field $d_0$, the class of "parallel curves" is characterized by a non-unique bilateral CPP. Konyukhov et al. argue that the contact angle~$\alpha$ employed in the criterion~\eqref{criterion_estimate} is not a suitable metric for assessing the uniqueness of the CPP since such "parallel curves", i.e. curves with non-unique CPP, can be constructed for arbitrary contact angles $\alpha$. In the following, we will show that this statement is not correct as long as $\mu_{max}:= R max(\bar{\kappa}) \ll 1$, which is typically fulfilled for beam models.\\

\begin{figure}[ht]
 \centering
   \subfigure[Two parallel beams]
   {
    \includegraphics[height=0.17\textwidth]{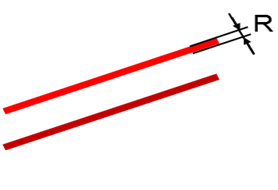}
    \label{fig:line_limitations_parallel}
   }
   \hspace{0.1\textwidth}
   \subfigure[Straight + circular beam]
   {
    \includegraphics[height=0.17\textwidth]{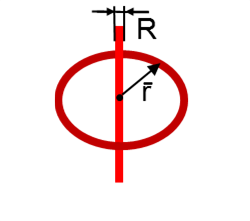}
    \label{fig:limitations_circle}
   }
   \hspace{0.1\textwidth}
   \subfigure[Straight + helical beam]
   {
    \includegraphics[height=0.20\textwidth]{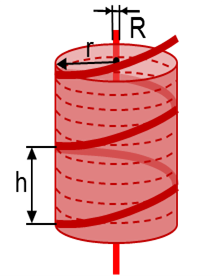}
    \label{fig:limitations_helix}
   }
  \caption{Contact of two beams: Geometrical configurations characterized by a constant distance function $d(s)=d_0$ (taken from~\cite{meier2015b}).}
  \label{fig:limitations}
\end{figure}

In order to determine the contact angle $\alpha$ between "parallel curves", the tangent vectors $\mb{t}_a(s)= d \mb{r}_a(s) / ds$ and $\mb{t}_b(s)= d \mb{r}_b(s) / ds$ of the two space curves are required, where the latter directly follows from the definition of "parallel curves" stated above. In a next step, Konyukhov et al. determine the angle between the tangent vectors according to $\cos(\alpha)=\mb{t}_a^T \mb{t}_b$. First, it has to be stated that this expression is not correct, but has to be replaced by $\cos(\alpha)=\mb{t}_a^T \mb{t}_b / ||\mb{t}_b||$, since in general only $\mb{t}_a(s)$, but not $\mb{t}_b(s)$, is a unit vector ($s$ is an arc-length parameter of $\mb{r}_a(s)$ but not of $\mb{r}_b(s)$), which can easily be verified. Second, Konyukhov et al. argue that "parallel curves" (with non-unique bilateral CPP) can easily be constructed even for the extreme case of orthogonal tangent vectors at the contact points, i.e. $\alpha=\pi/2$. Taking the expression for $\cos(\alpha)$ from above, a contact angle of $\alpha=\pi/2$ would mean $\cos(\alpha)\dot{=}0 \rightarrow  \mb{t}_a^T \mb{t}_b\dot{=}0$. Evaluating this expression according to the definition of "parallel curves" eventually yields $1+\bar{\kappa}_a(s)d_0\cos(\beta_a(s)) \dot{=}0$ (this criterion has been stated correctly by Konyukhov et al.~\cite{konyukhov2017} since the requirement $\cos(\alpha)\dot{=}0$ does not depend on the denominator of the expression $\cos(\alpha)=\mb{t}_a^T \mb{t}_b / ||\mb{t}_b||$, which was incorrect / not present in~\cite{konyukhov2017}). For arbitrary space curves, this condition can be fulfilled in general. However, as already investigated in the discussion of criterion~\eqref{criterion_general_a} above (see also~\ref{anhang:importance_of_assumptions} above), the fulfillment of this condition requires a minimal curvature of $\bar{\kappa}_a=1/d_0$ (which again corresponds to a configuration similar to Figure~\ref{fig:limitations_circle}). If beams are considered that are close enough such that contact can actually occur (i.e. $d_0 \leq 2R$, see assumption i)), the minimal curvature for which the condition $\cos(\alpha)\dot{=}0$ can be fulfilled is given by $\bar{\kappa}_a=1/(2R)$. Such a configuration, however, would violate assumption ii), which requires $\bar{\kappa}_a<1/(2R)$. It can be concluded that configurations with non-unique bilateral CPP and $\alpha=\pi/2$ can indeed occur for arbitrary space curves, but not in beam-to-beam contact scenarios fulfilling assumptions i) and ii). As already discussed above (see Section~\ref{central_criticism}), the application of geometrically nonlinear beam theories typically allows to state a maximal admissible ratio of curvature to cross-section radius even satisfying $\mu_{max}= \bar{\kappa}R \ll 1$. Since criterion~\eqref{criterion_estimate_b} for $\alpha_{min}$ is a function that increases monotonically with the maximally admissible curvature ratio $\mu_{max}$, it can be argued that small admissible curvature ratios $\mu_{max}\ll 1$ in turn lead to small values of the critical contact angle $\alpha_{min}$ below which configurations with non-unique bilateral CPP can occur.\\

This statement shall briefly be illustrated by a specific example of "parallel curves" (with constant distance $d_0=r$), as illustrated in Figure~\ref{fig:limitations_helix}: a straight beam is surrounded by a helicoidal beam with helix radius $r$ and helix slope $h$ (undeformable, circular cross-section with radius $R$ for both beams). For this geometry, the contact angle (i.e. the angle between the tangent vectors at a pair of closest points) can easily be derived as $\cos (\alpha)=h/\sqrt{r^2+h^2}$, and the curvature of the helicoidal beam centerline is given by $\bar{\kappa}=r/(r^2+h^2)$. Inserting the second relation into the first relation and considering only active contact configurations (limiting case $r=2R$, see again assumption i)), eventually yields: $\alpha(\mu)=\arccos (1/\sqrt{1+2\mu/(1-2\mu)})$ with $\mu=\bar{\kappa}R$. Between the two extreme cases $\mu=0$ (i.e. $h \rightarrow \infty$, see Figure~\ref{fig:line_limitations_parallel}) and $\mu=0.5$ (i.e. $h = 0$, see Figure~\ref{fig:limitations_circle}), the function $\alpha(\mu)$ is monotonically increasing (as the right-hand side of criterion~\eqref{criterion_estimate_b} is). For the contact of straight beams ($\mu=0$, see Figure~\ref{fig:line_limitations_parallel}), the well-known requirement $\alpha>\alpha_{min}=0$ is recovered by the criterion~\eqref{criterion_estimate_b}. Also the second extreme case is correctly represented by the criterion~\eqref{criterion_estimate_b}: As already discussed above, a critical contact angle of $\alpha_{min}=\pi/2$ can only occur for an impermissibly high curvature ratio $\mu=0.5$ (see Figure~\ref{fig:limitations_circle} for the case $\bar{r}=2R$), which is not compatible with assumption ii) and the limitations of typical geometrically nonlinear beam theories ($\mu_{max}\ll 1$). For example, a curvature ratio of $\mu=0.01$, which can be considered as reasonable upper bound for a typical beam model, yields an actual contact angle of $\alpha(\mu=0.01)\approx 8.1^{\circ}$ for the helix example~\ref{fig:limitations_helix}. On the other hand, the criterion~\eqref{criterion_estimate_b} guarantees a unique bilateral CPP for angles $\alpha>\alpha_{min}(\mu=0.01)\approx 11.5^{\circ}$. As expected, the worst case estimate correctly identifies the helicoidal geometry as configuration with non-unique bilateral CPP, i.e. $\alpha(\mu=0.01)<\alpha_{min}(\mu=0.01)$.\\

All in all, it can be concluded that it is indeed possible to generate examples with non-unique bilateral CPP even for beams contacting at an angle of $\alpha=\pi/2$ between the tangent vectors (see e.g. Figure~\ref{fig:limitations_circle} for the choice $\bar{r}=2R$). However, such a configuration would violate the basic assumption ii) in terms of centerline curvatures being too high. The restriction of typical structural beam theories to small curvatures $\mu_{max}\ll 1$, in turn, also leads to small values of the critical contact angle $\alpha_{min}$ below which active contact configurations with non-unique bilateral CPP can occur.

%
\subsection{Statement 5}
\label{statement_5}
%

In Section 7.2, Konyukhov et al.~\cite{konyukhov2017} analyze the dimension of the contact zone ocurring between two straight beams as function of the contact angle $\alpha$ and compare their results with the dimensions predicted by the Hertz contact theory. The Hertz contact theory predicts a scaling of the contact zone length $a_x$ (with $x$ representing the direction of the slave beam centerline) according to $a_x \sim 1/|\sin(\alpha)|$. In this context, Konyukhov et al.~\cite{konyukhov2017} state that criterion~\eqref{criterion_estimate} as derived in~\cite{meier2015b} "is not capturing" this behavior "at all". This argumentation, however, does not make sense since criterion~\eqref{criterion_estimate} is not related to the prediction of contact zone sizes in any way. In~\cite{meier2015c}, criterion~\eqref{criterion_estimate_b} has rather been employed to predict a proper interval of contact angles in which a smooth model transition from line-to-line contact (applied in the range of small contact angles) to point-to-point contact (applied in the range of large contact angles) is performed. In this context, two important aspects should be considered: First, it can easily be shown that the contact zone length predicted by the line-to-line contact (sub-)model for the contact of two straight beams also shows the correct scaling behavior, i.e. $a_x \sim 1/|\sin(\alpha)|$ (see equation (30) in~\cite{meier2015c}). Second, the transition from the line to the point contact model in the range of sufficiently large contact angles as proposed in~\cite{meier2015c} is motivated by considerable efficiency gains enabled by the latter type of model. Since the point contact model assumes a localized contact force, it is clear that the line contact model, which accounts for distributed contact forces and a finite contact zone size, represents a more accurate mechanical model in the range of small and moderate contact angles. However, it has to be noted that criterion~\eqref{criterion_estimate_b} only predicts a minimal contact angle $\alpha_{min}$, above which the point contact model is applicable at all (in terms of unique bilateral CPPs). Of course, in practice, the transition between line and point contact can also be performed at larger contact angles, which typically requires a compromise between model accuracy and computational efficiency. Nevertheless, in many practical applications (e.g. the modeling of complex networks of slender fibers), a reasonably accurate representation of the non-penetration condition and of net contact forces is of primary interest rather than an accurate resolution of contact pressure distributions on the beam surfaces. This often justifies the application of the point contact model already at comparatively small contact angles (above $\alpha_{min}$).

%
\bibliographystyle{plain}
\bibliography{beamreferences.bib}
%

\end{document}